\documentclass[12pt,preprint]{aastex}

\newcommand{\eb}{\begin{equation}}
\newcommand{\ee}{\end{equation}}

\def\3dot#1{\stackrel{...}{#1}}
\newcommand{\uas}{$\mu$as}

\newcommand{\kms}{km~s$^{-1}$}

\slugcomment{Version \today}

\shorttitle{Starspot jitter}
\shortauthors{Makarov et al.}

\begin{document}

\title{STARSPOT JITTER IN PHOTOMETRY, ASTROMETRY AND RADIAL VELOCITY MEASUREMENTS}

\author{V.V. Makarov\altaffilmark{1}, C.A. Beichman\altaffilmark{1}, J.H. Catanzarite \altaffilmark{2},
D.A. Fischer\altaffilmark{3}, J. Lebreton\altaffilmark{1}, F. Malbet\altaffilmark{1,4},
M. Shao\altaffilmark{2}}
\affil{$^1$NASA Exoplanet Science Institute, Caltech, \\ Pasadena, CA 91125}
\affil{$^2$JPL, Pasadena, CA 94550}
\affil{$^3$Department of Physics and Astronomy, San Francisco State University, San Francisco, CA 94132}
\affil{$^4$Centre National de la Recherche Scientifique, Paris, France}
\email{vvm@caltech.edu}

\begin{abstract}
Analytical relations are derived for the amplitude of astrometric,
photometric and radial velocity perturbations caused by a single rotating spot.
The relative power of the star spot jitter is estimated and compared with
the available data for $\kappa^1$ Ceti and HD 166435, as well as
with numerical simulations for $\kappa^1$ Ceti and the Sun. A Sun-like star
inclined at $i=90\degr$ at 10 pc is predicted to have a RMS jitter of
$0.087$ \uas\ in its astrometric position along the equator, and
$0.38$ m s$^{-1}$ in radial velocities. If the presence of spots due to stellar activity 
is the ultimate
limiting factor for planet detection, the sensitivity of SIM Lite to Earth-like 
planets in habitable zones
is about an order of magnitude higher that the sensitivity of prospective
ultra-precise radial velocity observations of nearby stars.
\end{abstract}


\section{Introduction}
With the anticipated launch of the SIM Lite mission in the near future,
we are embarking on a long and exciting journey of exoplanet
detection by astrometric means. One of the main goals of this mission is the detection 
of habitable
Earth-like planets around nearby stars \citep{unw}. To date, most of the
exoplanet discoveries have been made by the Doppler-shift technique, while
the astrometric method has been limited to the use of the FGS on
the Hubble Space Telescope \citep{ben} and to ground-based CCD observations of
low-mass stars with giant, super-Jupiter,  planetary companions \citep{sha}. 
In achieving the strategic goal of confident
detection of rocky, Earth-sized planets in the habitable zone, the prospective
astrometric and spectroscopic ultra-precise measurements will
encounter a number of limitations of technical and astrophysical nature. 

For the Doppler-shift technique, many of 
these limitations will be dealt with by
further improvement in the instrumentation or refinement of the observational
procedure \citep{may}. However, the presence of astrophysical noise due to stellar
magnetic activity emerges as the ultimate bound on the sensitivity of planet
detection techniques,
and the only remedy suggested thus far, is selection of particularly inactive,
slowly rotating stars. Indeed, a very small fraction of stars in the
high-precision HARPS program of exoplanet search exhibit radial velocity (RV)
scatter of less than 0.5 m s$^{-1}$. Although this type of variability is probably 
driven by the rotation
of bright and dark structures on the surface (star spots and plage areas), the frequency power
spectrum of such perturbations can be fairly flat, extending to frequencies much higher
or lower than the rotation, as shown by \citet{cat} for the Sun. Arguments have been
presented \citep[e.g., ][]{eri} that star spots also result in very large astrometric noise
of $\sim 10$ $\mu$AU, which should thwart discovery of habitable Earth
analogs. The aim of this paper is to quantify the effects of rotating spots in astrometric
photometric and RV measurements more accurately, taking into account the limb darkening,
geometric projection and differential rotation, and to assess the expected vulnerability
of the RV and astrometric methods to such perturbations. We do this by direct analysis
as well as by numerical simulation, and support our findings by the data for the Sun
and two rapidly rotating stars.

\section{Perturbations from a single spot}
\label{jit.sec}
We consider a single circular spot whose instantaneous position on the surface in the stellar
reference frame is given by longitude $l$ and latitude $b$, which are the angles from the direction
to the observer and from the equator, respectively. The projected area of the spot is $\pi r \cos C$,
where $C$ is the central angle between the direction to the observer and the center of the spot,
and $r$ is the radius of the spot in radians, $r<< 1$.
The position vector of the spot in the local sky triad $\{\Re, E, N\}$ is ${\bf s} =
[s_1, s_2, s_3]^{\rm T} = [-\cos l\,\cos b\, \sin i - \sin b\,\cos i,\, -\sin l\,\cos b,
 -\cos l\,\cos b\,\cos i+\sin b\,\sin i]^{\rm T}$. $N$ is north, $E$ is
east, and $\Re$ is the line-of-sight directions in this right-handed triad.
The velocity vector of the spot is ${\bf V} =
[v_1, v_2, v_3]^{\rm T} = [\sin l\, \sin i ,\, -\cos l,\,
\sin l\, \cos i]^{\rm T}V_b$, where the differential rotation velocity
\eb
V_b=\frac{2\pi R\,\cos\,b}{P_{\rm rot}(b)}\approx V_{\rm eq}\,\cos\,b\,(1-0.19\sin^2 b),
\ee
where $V_{\rm eq}$ is the equatorial rotation velocity. This equation assumes the
differential relation for the Sun derived from sunspot latitudes and periods \citep{new,kit}.
We also assume that the spot's contrast with respect to the local surface brightness is
fixed at $f_{\rm s}$. The asymmetry in the distribution of surface brightness due to the
spot results in certain perturbations in the integrated flux, photocenter and radial velocity
of the stellar disk. Assuming the limb darkening for the Sun at $\lambda=550$ nm
\eb
\frac{I(C)}{I(0)}=0.30+0.93 \cos C-0.23\cos^2 C,
\ee
the integrated flux from the stellar disk is $0.905\pi I(0)$, the following relations are
obtained for the amplitudes of perturbation
\begin{eqnarray}
\label{delta.eq}
\frac{\Delta F}{F} &=&-(1-f_{\rm s})\phantom{\, s_2}\, r^2\, \frac{I(C)\cos C}{0.905\, I(0)}\\
\Delta x &=&-(1-f_{\rm s})\, s_2\,r^2\, R\,\frac{I(C)\cos C}{0.905\, I(0)}\nonumber\\ 
\Delta y &=&-(1-f_{\rm s})\, s_3\,r^2\, R\,\frac{I(C)\cos C}{0.905\, I(0)}\nonumber\\ 
\Delta V_{\rm R} &=&-(1-f_{\rm s})\, v_1\,r^2\, V_{\rm eq}\,\cos\,b\,(1-0.19\sin^2 b)\;\frac{I(C)\cos C}{0.905\, I(0)} \nonumber
\end{eqnarray}
where $R$ is the apparent radius of the star. These expressions describe the modulation of
the flux, photocenter, and radial velocity of the star due to the motion of a spot.
For the Sun, $V_{\rm eq}=2$ \kms\ and $R_\odot=4650$ $\mu$AU. We estimate a relative flux
variability of the Sun of RMS$(\Delta F/F)=3.24\cdot 10^{-4}$ after subtracting a 10-yr
period solar cycle light curve from the solar irradiance PMOD data \citep{fro}. Therefore,
the sunspot-related jitter is not greater than $\Delta m\,R$ ($1.5$ $\mu$AU for
the Sun) in position and than $\Delta m\,V_{\rm eq}$ ($0.65$ m s$^{-1}$  for
the Sun) in radial velocities, where  $\Delta m$ is the characteristic magnitude
jitter. The astrometric perturbation decreases with distance to a typical value
of $0.15$ \uas~ for the Sun at $D=10$ pc.

\begin{figure}[htbp]
\plotone{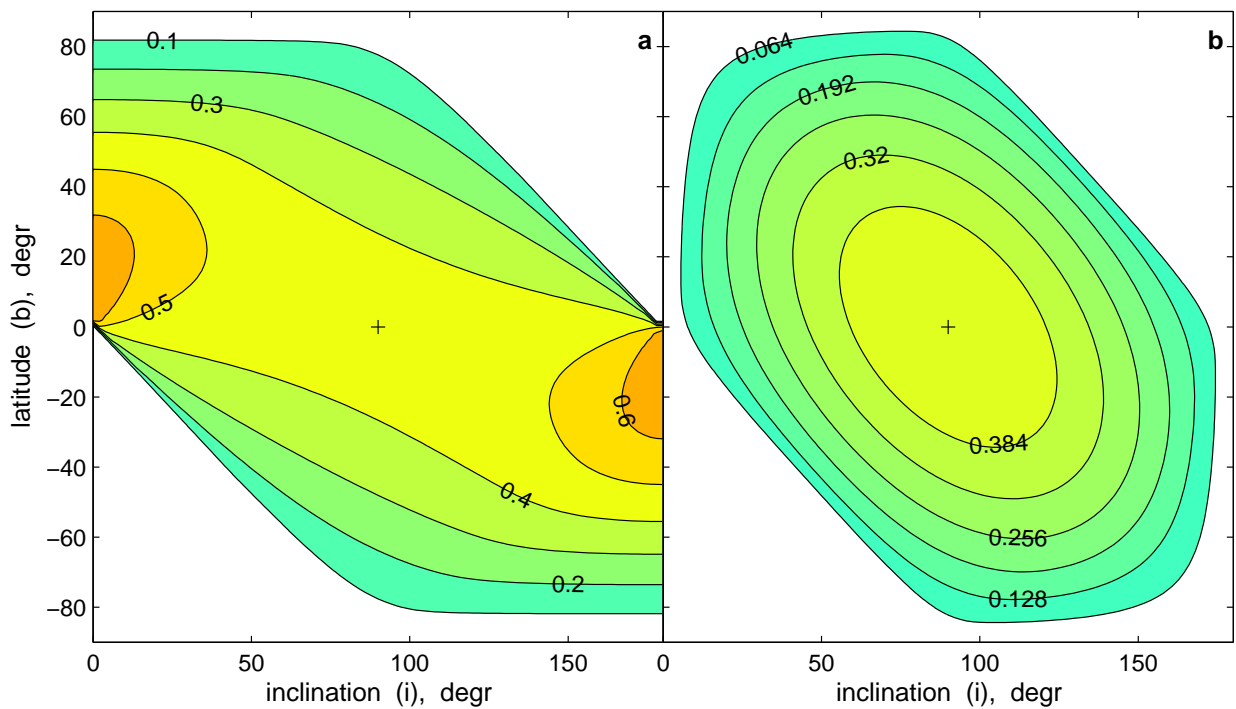}
\caption{The magnitudes of relative RMS perturbations from a 
single star spot: a) ratio RMS($\Delta x$)$/$
RMS($\Delta F/F$) in units of apparent stellar radius $R$; b) ratio RMS($\Delta V_{\rm R}$)$/$
RMS($\Delta F/F$) in units of $V_{\rm eq}(1-0.19\sin^2 b)$. In both cases the values
at $i=90\degr$, $b=0\degr$ are 0.448.} 
  \label{RMS.fig}
\end{figure}

Eqs.~\ref{delta.eq} can be used for numerical simulation of star spot perturbations in a computationally
efficient way. They can also be integrated in quadratures to estimate the power (root-mean-square,
RMS) of the jitter. This results in fairly tedious series in powers of sines and cosines of
$b$ and $i$, which we do not give here for brevity. Some of the results are represented in graphical
form in Fig.~\ref{RMS.fig}. It should be noted that the ratios of RMS values in flux and each
of the remaining parameters can only be directly computed for a single spot on the surface.
Using this figure, and the RMS jitter for the Sun at $i=90\degr$ and $b=0\degr$ are
${\rm RMS}(\Delta x)=0.87$ $\mu$AU and ${\rm RMS}(\Delta V_{\rm R})=0.38$ m s$^{-1}$.

The geometric projection factors $s_2$ and $v_1\,\cos\,b$ differ only by a constant factor $-\sin\,i$.
Therefore, the ratio of the perturbation amplitudes, as well as perturbation RMS in $\Delta x$ and
in $\Delta V_{\rm R}$ is constant for a single spot:
\eb
\frac{{\rm RMS}(\Delta V_{\rm R})}{{\rm RMS}(\Delta x)}=\frac{V_{\rm eq}\,\sin\,i}{R}(1-0.19\sin^2b).
\ee
Using the values for the Sun, the approximate scaling relation is
\eb
\frac{{\rm RMS}(\Delta V_{\rm R})}{{\rm RMS}(\Delta x)}\approx 0.43\,\sin\,i\,\left[ \frac{D}{{\rm 1~pc}}
\right] \left[ \frac{P_{{\rm rot},\odot}}{P_{\rm rot}}\right],
\label{vrdx.eq}
\ee
in units m s$^{-1}$ \uas$^{-1}$, where $P_{{\rm rot},\odot}=24.47$ d is the sidereal rotation
period of the Sun at the equator. Using the Carrington period of $25.38$ d instead will
to some extent account for the distribution of sunspots in latitude, and allows one to
ignore the differential rotation factor to first-order approximation. 

The data in Fig.~\ref{RMS.fig} and Eq.~\ref{vrdx.eq} can be used to estimate the relative
magnitude of starspot jitter in astrometry and RV measurements. Eqs.~\ref{delta.eq}
provide an efficient and direct way of simulating these perturbations for any 
configuration of spots.

\section{Comparison with observations}
\label{known.sec}
\subsection{HD 166435}
The star HD 166435 is a solar-type dwarf at 25 pc without conspicuous signs of 
chromospheric or coronal activity,
which nonetheless exhibits large and correlated variations in brightness, radial velocity,
and CaII H and K lines \citep{que}. Strong evidence is presented in that paper that these
periodic variations are caused by a photospheric spot or a group of spots, including
analysis of spectral line bisectors. Some confusion with a possible short-period giant
planet resulted from the conspicuously large amplitude of the RV variation ($\simeq 200$
m s$^{-1}$ peak-to-peak) and the stable phase on a time scale of 30 days. Queloz et al. obtained
a projected rotation velocity of $v\,\sin\,i = 7.6\pm 0.5$ \kms\, and a period of 3.7987 d. They
estimated a $i=30\degr$ from these data. The shape of the variability curves is consistent
with a single dominating spot rotating with this period.

Ignoring the differential rotation term (which is
probably small for fast-rotating stars) and 
setting $\Delta V_{\rm R}$ to 200 m s$^{-1}$ from Queloz et al.'s Fig. 8, or
166 m s$^{-1}$ from their text, and $\Delta m$ to $0.1$ mag or $0.07$ mag,
we derive from Eqs.~\ref{delta.eq} $\cos\,b=0.286$, $b=73\degr$, or $\cos\,b=0.339$, $b=70\degr$, respectively.
This latitude is ambiguous with respect to sign, the possible combinations
being $i=30\degr$, $b=70\degr$ (counterclockwise rotation), or $i=150\degr$, $b=-70\degr$ 
(clockwise rotation). In any case, the center of the spot is close to the pole, and because of the
small inclination, circles close to the middle of the visible stellar disk. This is
fully consistent with the conclusion by \citet{que}, which they draw from the
smoothness of the variability curves. Finally, using the above estimates for $i$ and $b$,
we compute the light curve, which is indeed a smooth sinusoid-like function of time,
with a peak-to-peak amplitude of $0.51(1-f_{\rm s})r^2$. For a large spot area, the factor
$r^2$ is interpreted as the fraction of the observed hemisphere covered by the spot.
The characteristic contrast ratio of sunspots is 0.2 in the optical passband, which
corresponds to an effective temperature of $\sim 4400$ K for the spotted photosphere
\citep{lan}.
This yields a striking number for the area of the spot, $r^2=0.23$, or an angular radius
$\rho=40\degr$. This feature appears to be a dark sea
engulfing the pole. \citet{que} find that the phase of the RV variation is confined to a
$\pm0.1$ interval (their Fig.~4), which indicates that
the feature is stable on a time scale of 2 years, but with considerable 
internal variations of brightness.

\subsection{$\kappa^1$ Ceti}
The available data for this star relevant to this study include the precision
photometric series from {\it MOST} and 44 individual RV measurements spread over
some 20 years, provided by one of us (DF). The {\it MOST} data sets have been
carefully analyzed in other papers \citep{ruc, wal,bia}, with a firm conclusion
that its light curve can be well modeled with a set of two or three dark spots.
The periods of rotation range between 8.3 and 9.3 days. The observed peak-to-peak
amplitudes in flux are roughly $0.05$ in 2003, but only $\approx 0.02$ in 2004
and 2005. The projected speed of rotation measured by \citep{val}, $v\,\sin\,i=5.2\pm
0.5$ \kms\ implies, from Eq.~\ref{delta.eq}, a RV perturbation 
$\leq 260$ m s$^{-1}$ in 2003 and $\leq 120$ m s$^{-1}$ in 2004
and 2005. Due to the sparsity of the RV data, we can not match them directly with
the intervals of MOST observations, but we can estimate the amplitude of RV variation
over two decades. With the smallest observed value of $-68.56\pm 5.01$ m s$^{-1}$, and the largest
$+42.50\pm 9.88$ m s$^{-1}$, the amplitude is close to the single-spot estimate for the more
quiescent periods, but is much smaller than the prediction for 2003. 
Independent observations by \citet{wal95} during 1982--1992 also indicated a peak-to-peak
amplitude of $\sim 100$ m s$^{-1}$.
One possible explanation is that the spots usually reside at high
latitudes, which reduces the relative variation in RV because of the $\cos\, b$
factor. The smaller RV variability may also be related to the fact,
that two or three spot groups are present on the surface at a time, rather than
one single spot. Eqs.~\ref{delta.eq} can not be simply scaled for the case of multiple
spots, because the $\Delta V_{\rm R}$ and $\Delta x$ function of time are symmetric
around the central meridian ($C=0$). As a result, two spots well separated in longitude
can counterbalance each other, reducing or nearly canceling the net perturbation.
Indeed, the detailed modeling by \citet{wal} indicate that large spots frequently
occur in the near-polar regions of $\kappa^1$ Ceti, and that two or three coexisting
spots are spread in longitude, rather than grouped in a confined active area.

\begin{figure}[htbp]
\plotone{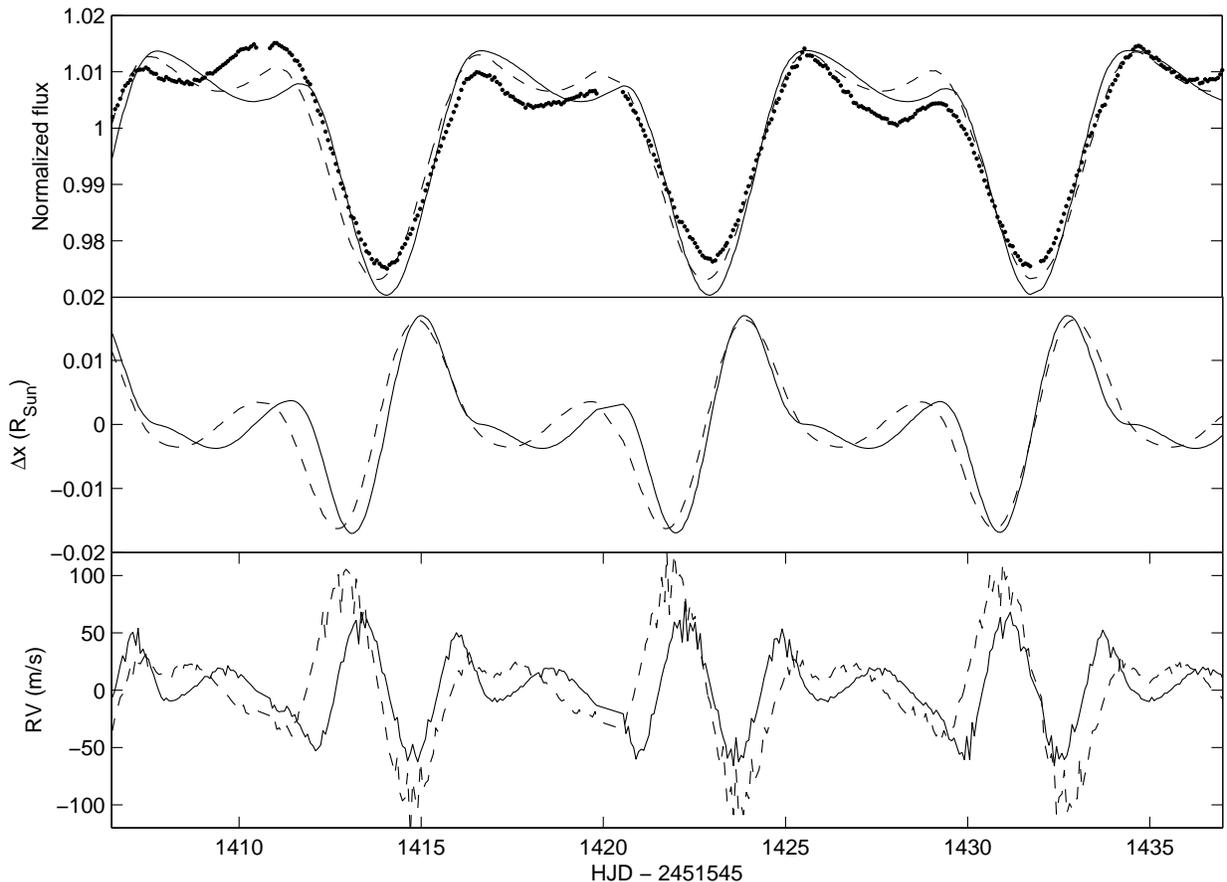}
\caption{Numerical simulations of variation in relative flux, equatorial shift
of the photocenter, and radial velocity of $\kappa^1$ Ceti caused by two rotating
spots, corresponding to the first segment of observations with {\it MOST}
in 2003. Our prediction is drawn with a solid line, \citet{wal} results with
dashed line, and {\it MOST} data with asterisks.} 
  \label{curve.fig}
\end{figure}
Learning more about the properties of spots on this star requires more accurate
computation. One of us (JL) created a code to model differentially rotating star spots and
the resulting perturbations of the observable parameters by pixelization of the
stellar surface and integration over the visible hemisphere. The free model parameters
were optimized on the photometric series of $\kappa^1$ Ceti, essentially
repeating the study by \citet{wal}, but also extending it to astrometric and radial
velocity predictions. A detailed description of this model will be published elsewhere,
here we only discuss some of the results relevant for this paper. Fig.~\ref{curve.fig}
shows the simulated perturbations from our model, along with the expected
perturbations from the model by Walker et al., and the actual light curve from
{\it MOST} for the segment of 2003. With only two spots in both cases, the
goodness of photometric fit is similar with the two models, but our simulation predicts a smaller
amplitude of RV variation ($\sim 110$ m s$^{-1}$), probably because of a more symmetric 
configuration of spots. This prediction is in fact consistent with the available
RV data. The ratio of simulated variations in $\Delta V_{\rm R}$ and $\Delta x$
is consistent with Eq.~\ref{vrdx.eq}.

\section{The Sun}
\label{sun.sec}
We estimated in \S~\ref{jit.sec} that the expected RMS jitter in RV for the Sun is
$0.38$ m s$^{-1}$. The half-amplitude of the reflex motion caused by the
Earth orbiting the Sun is slightly less than $0.1$ m s$^{-1}$. Does this result imply
that detection of Earth-like planets in the habitable zone of Sun-like
stars is impossible? This brings up the more subtle issue of the
spectral power distribution of starspot jitter. The characteristic period
of star spot rotation is 1 month for Solar-type stars, but the orbital
periods of habitable planets are of order 1 year. It is therefore not obvious
without more detailed analysis that the signal-to-noise ratio will be too small
for a confident detection in the frequency domain of interest.

One of us (JC) performed extensive Monte-Carlo simulations for the Sun,
which are described in more detail in \citep{cat}. Sunspot groups are
generated randomly through a Poisson process, with probability distributions
consistent with the current data. The main purpose of this simulation
is to faithfully reproduce the power spectrum of the solar irradiance data
on a time scale of 30 years. The average number of sunspot groups and the
dispersion of lifetimes was adjusted in such a way that the predicted and
the observed spectral power of $\Delta F/F$ match closely in the frequency
range $10^{-7}$ to $4\times 10^{-8}$ Hz. This gives some assurance that the
model predictions can be accurately made for the power spectra of the
astrometric and RV jitters. Typical RMS values of jitter predicted by
the numerical simulation at $i=90\degr$ are: $0.39$ m s$^{-1}$ for RV, $0.97$ $\mu$AU in $x$,
and $0.52$ $\mu$AU in $y$, quite close to the analytical estimates in \S~\ref{jit.sec}.

\begin{figure}[htbp]
\plottwo{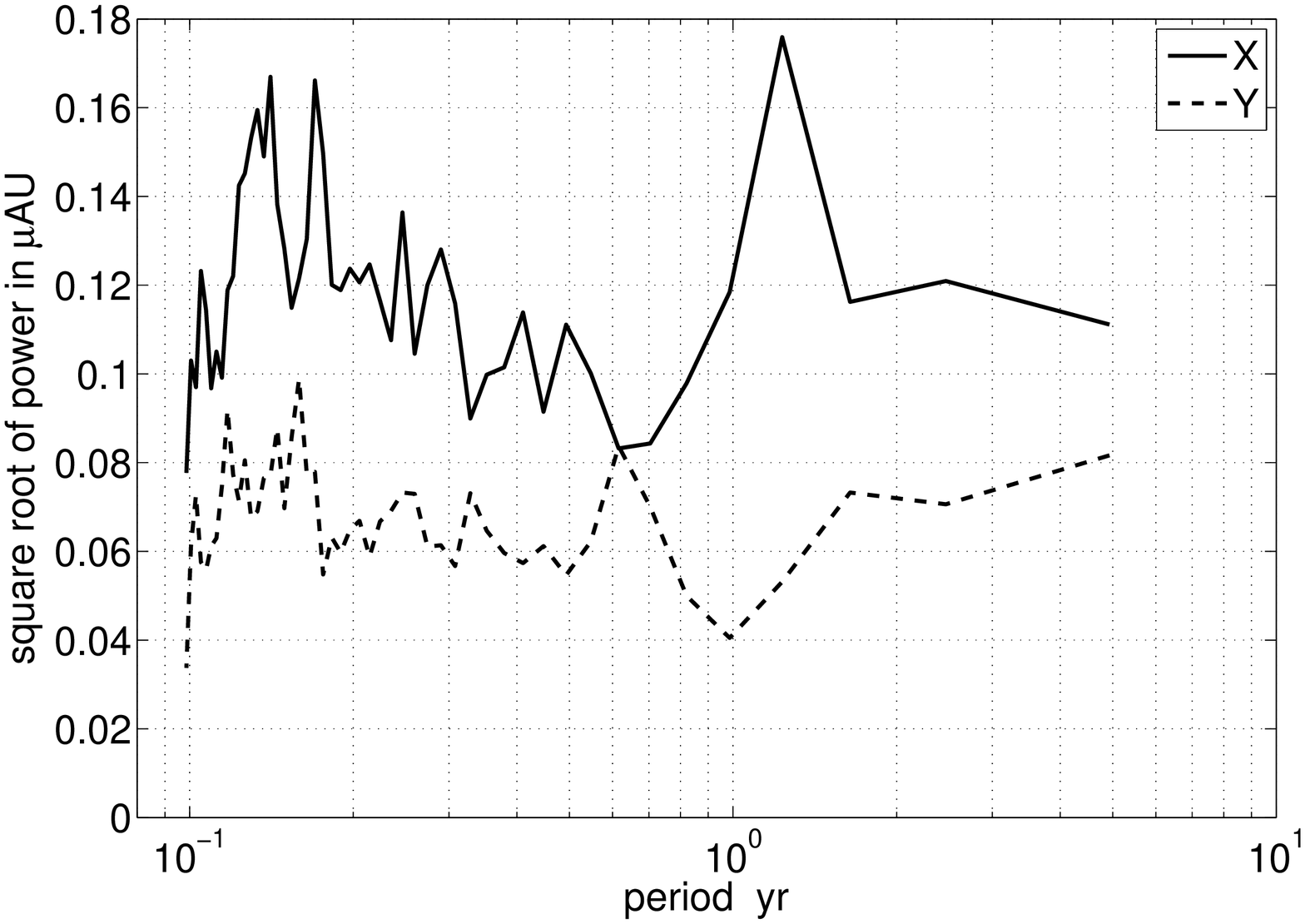}{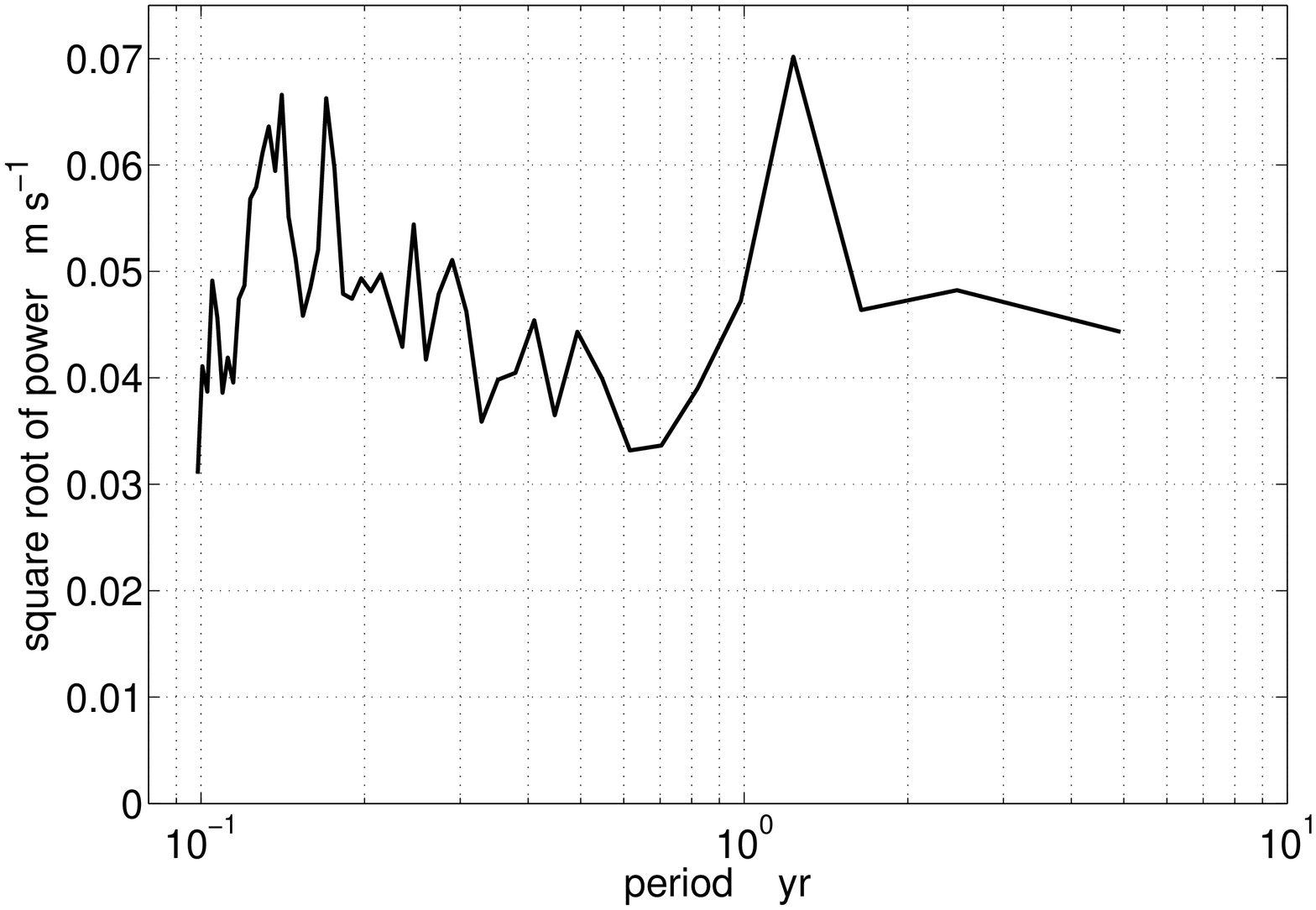}
\caption{Square root of the spectral power of simulated astrometric (left plot)
and radial velocity (right plot) perturbations
of the Sun seen at inclination $90\degr$. The X axis is aligned with the equator,
and the Y with the rotation axis.} 
  \label{sunXY.fig}
\end{figure}

Fig.~\ref{sunXY.fig} shows the square root of power in 100 astrometric and
radial velocity observations
over 5 years simulated for the Sun
at $i=90\degr$.  The equator is expected
to be coplanar with the orbital plane, hence, the exoplanet signal will
be present only in the $x$-measurements. The power of spot-related jitter spreads
far and wide from the rotation period of 25 days. It peaks between periods
of 1 and 2 years, reaching almost $0.18$ $\mu$AU in astrometry. In the worst case, exoplanets
with signatures of $0.63$ $\mu$AU or greater can be confidently discovered with
SNR$>3.5$ by the astrometric method. The spectrum of the simulated RV variations
is practically identical to the spectrum of $x$-jitter,
as predicted in \S~\ref{jit.sec}. The peak value is $0.07$ m s$^{-1}$. The corresponding
semiamplitude of exoplanet signature detectable at SNR$=3.5$ is $0.25$ m s$^{-1}$.

\section{Conclusions}
\label{con.sec}
Our results for the Sun are in good agreement with the approximate relations by
\citet{eri}, who estimated a positional standard deviation of 0.7 $\mu$AU.
At the same time, their conclusion that "for most spectral types the astrometric
jitter is expected to be of the order of 10 micro-AU or greater" is misleading,
because it is largely based on overestimated values of photometric variability
from ground-based observations, and it does not differentiate the luminosity classes of
giants and dwarfs. It can not be concluded that the Sun
is exceptionally inactive compared to its peers just because the ultra-precise
solar irradiance data, such as PMOD or SOHO, reveal a much smaller scatter than
the inferior photometric data for other stars. We investigated the indices of
chromospheric activity ($\log R^\prime_{\rm HK}$) and available rotation periods
for some 80 SIM targets, and found that half of them should rotate with the same
rate as the Sun, or slower.

\citet{hal} presented a detailed study of variability of solar-type stars and
its relation to the index of chromospheric activity $\log R'_{\rm HK}$, based
on 14 years of photometric and spectroscopic observations. They found that the
Sun at $\log R'_{\rm HK}=-4.96$ is not more variable than its F-G peers at the low
end of the activity distribution. Given that most Solar-type stars exhibit
similar low levels of chromospheric activity \citep{gra}, we expect that finding
stars with levels of jitter similar to, or lower than the Sun, should not be a
problem. Low-jitter, stable stars are common and plentiful, which augurs well
for the prospects of finding small, rocky planets with {\it Kepler} \citep{bat}.

\begin{deluxetable}{lrrr}
\tabletypesize{\scriptsize}
\tablecaption{Observable signals and star spot jitters for an Earth-like
planet orbiting a typical dwarf star at 10 pc. \label{snr.tab}}
\tablewidth{0pt}
\startdata
  Star type \dotfill & Sun & F5V & K5V \\
  Rotation period, d \dotfill & 25.4 & 18 & 30 \\
  Astrometric signal, \uas\ \dotfill & 0.30 & 0.23 & 0.45 \\
 RV signal, m s$^{-1}$ \dotfill & 0.089 & 0.078 & 0.109 \\
Astrometric jitter, \uas\ \dotfill & 0.087 & 0.113 & 0.063\\
RV jitter, m s$^{-1}$ \dotfill & 0.38 & 0.69 & 0.23  \\
Astrometric SNR \dotfill & 3.4 & 2.0 & 7.1 \\
RV SNR \dotfill & 0.23 & 0.11 & 0.47 \\
\enddata
\end{deluxetable}
 
The SIM Lite Astrometric Observatory (formerly known as the Space
Interferometry Mission) will achieve a single-measurement accuracy
of 1 \uas\ or better in the differential regime of observation \citep{unw, sim}.
Several previous studies have addressed SIM's exoplanet detection and orbital
characterization capabilities \citep[][and references therein]{catpa}. The
"Tier 1" program includes $\sim 60$ nearby stars for which the astrometric signature
of a terrestrial habitable planet is large enough to be confidently measured by
SIM. The recently completed double-blind test \citep{tra} demonstrated that
Earth-like planets around nearby stars can be discovered and measured even
in complex planetary systems. In this paper, we are concerned with the more
general question of the ultimate limit to planet detectability set by the activity-related
jitter. Assuming that the
instrumentation progresses to levels where observational noise becomes insignificant, 
which technique holds the best prospects for detection
of habitable Earth-like planets? Table \ref{snr.tab} summarizes the expected
RMS jitter and the exoplanet signal for three typical nearby stars. In all cases,
the solar spot filling factor ($r^2$) is assumed. The signal-to-noise ratios (SNR) per 
observation include only the star spot jitter. Note that the astrometric SNR in this case is
independent of the distance to the host star, because both the signal and the
star spot jitter are inversely proportional to distance. As a measure of the relative sensitivity
of the two methods, the ratio of the SNR values for a given star is
independent of the planet mass or the filling factor to first-order approximation.
The physical radius of the star and the period of rotation are the two
parameters with the largest impact on the relative sensitivity, but their combined
effect is rather modest for normal stars, as is seen in the Table.
Therefore, in the ultimate limit of exoplanet detection defined by intrinsic
astrophysical perturbations, the
astrometric method is at least an order of magnitude more sensitive than the
Doppler technique for most nearby solar-type stars.

\acknowledgments
The authors thank G. Walker for his detailed and helpful review.
The research described in this paper was in part carried out at the Jet Propulsion 
Laboratory, California Institute of Technology, under a contract with the National 
Aeronautics and Space Administration. This research has made use of the NStED database,
maintained at NExScI, Pasadena, USA.

\end{document}